\documentclass{elsart}
\newlength{\defaultparindent}
\newenvironment{Default Paragraph Font}{}{}
\usepackage{epsfig}
\usepackage{graphics}                    
\date{today}

\begin{document}

\begin{frontmatter}

\title{Theoretical Fluctuations of Conductance in Stretched Monatomic Nanowires}

\author[UFR]{Fabien Picaud} ,
\author[UFR]{Vincent Pouthier,} ,
\author[UFR]{Claude Girardet} ,
\author[SISSA,INFM,ICTP]{and Erio Tosatti}
\address[UFR]{Laboratoire de Physique Moleculaire. UMR CNRS 6624. Facult\'{e} des sciences, La Bouloie, Universit\'{e} de Franche-Comt\'{e}, 25030 Besan\c{c}on (France)}
\address[SISSA]{SISSA Via Beirut 2/4, 34014 Trieste (Italy)}
\address[INFM]{INFM Democritos National Simulation Center, and INFM,
  Unit\`a  SIIS, Via Beirut 2/4, 34014 Trieste
(Italy)}
\address[ICTP]{ICTP, Strada Costiera 11, 34014 Trieste (Italy)}

\maketitle

\begin{abstract}
Recent experiments showed that the last, single channel conductance step
in monatomic gold contacts exhibits
significant fluctuations as a function of stretching.
From simulations of a stretched gold nanowire 
linked to deformable tips, we determine the distribution of the bond lengths 
between atoms forming the nanocontact and analyze its influence on the electronic 
conductance within a simplified single channel approach. We show that the 
inhomogeneous distribution of bond lengths can explain the occurrence and the
5\% magnitude of conductance fluctuations below the quantum conductance unit $g_{o}= 2e^2/h$.


\end{abstract}
\end{frontmatter}

\section{Introduction}

Ballistic electronic conductance in short metallic nanowires and nanocontacts
between tips has been widely analyzed both experimentally and theoretically in the last
decade. The transport properties of the nanocontacts are known to
be influenced by the geometrical structure of the contacts and by the
electronic confinement \cite{NAKAMURA}. Several experimental \cite{RUBIO,LANDMAN} 
and theoretical \cite{BRANDBYGE1,TODOROV,TORRES,BARNETT,BRANDBYGE2,STAFFORD} 
papers discussed the interplay between geometry and conductance.

Mechanical properties of atomic contacts were accessed experimentally by scanning
tunneling and atomic force microscopy, (STM/AFM) \cite{RUBIO} and theoretically 
mostly by molecular dynamics simulations \cite{LANDMAN,TODOROV,STAFFORD,SORENSEN}. 
It was shown \cite{YANSON,OHNISHI} that it is possible to pull stable monatomic gold 
wires, up to 7 atoms in length between two gold electrodes, with a conductance 
close to $g_{o}=2e^{2}/h$. More recently, STM supplemented with a force sensor was
used \cite{RUBIO-BOLLINGER} to study the mechanical response under stretching
of a low temperature (T = 4K) chain of single Au atoms. \it{Ab initio} \normalfont 
calculations of the breaking strength and of other mechanical
properties of the nanowires demonstrated \cite{RUBIO-BOLLINGER} a very considerable 
strength of bonds at these low coordinations relative to the high coordinations of
the bulk metal.

Transport of electrons in metal nanowires was also analyzed \cite{RUBIO} 
and abrupt conductance jumps were associated with atomic
rearrangements in the wire. STM tips were used \cite{OHNISHI} to generate
gold nanowires \itshape{in situ} \normalfont in a high resolution transmission electronic
microscope while recording the conductance. These results revealed the
existence of suspended gold atom chains with a conductance equal to $g_{o}$, while
another study \cite{RODRIGUES1} addressed the correlation between gold
nanowire structure and the quantum conductance behavior within the same
technique.  

Theoretical approaches devoted to the understanding of the relation between
electronic transport and the geometrical structure of metallic nanocontacts 
has been extensively developed.\ Tight binding models
were proposed to analyze the conductance of gold wires in terms of electron
standing waves due to the interference of electronic waves reflected at the
extremities of the atomic constrictions \cite{EMBERLY} or to determine the
transmitting channels in the atomic chain in terms of s,p,d orbitals \cite{BRANDBYGE3}.
The conductance of linear chains of 4 Au atoms suspended
between two jellium electrodes was calculated \itshape{vs} \normalfont the distance between the
electrodes within the \itshape{ab initio} \normalfont local spin density functional (LSDF)
approach. The conductance was determined using the recursion transfer
matrix \cite{OKAMOTO} and shown to decrease from 1 g$_{o}$ as the chain was
stretched, similar to a Peierls transition.

More recently, the electronic transport in free standing gold
atomic chains of up to 7 atoms in length was studied at 4K
\cite{AGRAIT}. All along during the pulling of the monatomic nanowire,
the behavior of conductance was characterized by two features. The first
was a $5\%$ - level fluctuation as a function of stretching, or between one pull
and another. The second was a $1\%$-level conductance drop when the voltage
across the nanocontacts exceeded about 15 meV. The latter was interpreted 
as a dissipative effect, quite similar to losses observed in point contact
spectroscopy, this time due inelastic scattering of electrons
inside the nanowire, where longitudinal vibrations can be excited 
via electron-phonon interaction. The identification of the loss peak with
a phonon inside the nanowire could be made thanks to the large 
phonon softening and  strengthening of the loss intensity which is observed
upon stretching.

That interpretation  was directly supported by a recent DFT calculation
accompanied by a molecular dynamics simulation\cite{PICAUDTOS}. There it 
was also shown that only about half the total stretch magnitude is absorbed by
the nanowire itself, the other half being absorbed by the tip-wire
junctions. The strengthening of the longitudinal phonon loss intensity in the stretched
nanowire was also theoretically addressed and found to be directly related to the 
stretching-induced softening through perturbation theory\cite{STELLA}.

In this paper, we provide a more specific rationalization to the zero-voltage
ballistic conductance fluctuations around g$_{o}$. We do that by calculating 
explicitly, albeit approximately, the conductance in a nanocontact. To
that end we first built a mechanical model to represent the gold
nanowire suspended between two tips. Based on that, and on a one dimensional 
bond model between atoms of the wire connected to two infinite reservoirs, we 
calculated the wire conductance during the course of its stretching, as
described by the mechanical model. Focusing on the distribution of 
interatomic distances in the gold stretched wire, we analysed the conductance 
by means of a Green's function operator formalism, that takes into account the 
changes in the hopping integrals with the distance between atoms.

\section{Theoretical model}

\subsection{Model for stretched monatomic nanowire}

In order to model the structural properties of a gold wire stretched between two
tips, we use a semi-empirical effective potential, which is in turn based on tight binding
theory in the second moment approximation (SMA). The potential experienced
by the n$^{th}$ gold atom and due to the other gold atoms m separated by a
distance $r_{nm}$ is written as \cite{GAMBA}:

\begin{equation}
V_{n}=\lambda \sum_{m}e^{-p\left( \frac{r_{nm}}{r_{0}}-1\right) }-\epsilon
\left( \sum_{m}e^{-2q\left( \frac{r_{nm}}{r_{0}}-1\right) }\right) ^{\alpha }
\label{EQUA1}
\end{equation}

All the parameters entering in the above expression are calculated by
fitting the experimental bulk and surface properties of gold. Their values
are $\lambda =0.4086$ $eV$, $p=8.5624$, $\epsilon =1.6332$ $eV$, $q=3.6586$, $%
\alpha =0.6666$, $r_{0}=2.88$ \AA\ and we consider a cutoff function for distances
$r_{nm}>r_{c})$ where $r_{c}$ is the second nearest neighbor distance \cite{PICAUDTOS,GAMBA}.
This potential was used to determine the mechanical properties
(equilibrium spacing and cohesive energy) of a gold wire wedged between two
gold tips. It was found to give results in close agreement with those
calculated using a more accurate DFT approach \cite{TORRES,PICAUDTOS}.

The system is formed by some atoms which are allowed to move and other atoms
that are fixed, the latter mimicking the connecting tips.
The moving atoms are those forming the nanowire (we chose 7 atoms at the
beginning of the simulation) and the junction forming a part of the two tips.
Each tip contains 13 moving atoms arranged in a rectangular (110) lattice.
This moving system is blocked on each side by two rectangular planes of fixed gold atoms, 
forming the back sides of the tips, and the distance between the inner fixed planes
is noted $L$ (distance between the two black ball planes represented in Fig. 2).

As L is increased and the planes move
apart, they stretch the nanowire and the tip-wire junctions in between.  We 
simply optimize the total energy of the system submitted to a gradually 
increasing stress, without the need of thermal dynamics (the temperature of the experiments is
very low ($4K$)). All the moving atoms are free to find
their equilibrium positions, at each stretching step.

Conductance calculations are then carried out at each step of the stretching simulation.
We verify first that the geometry of the tips and of
the wire-tip junctions is free to evolve, and thus to influence the evolution of the
wire during stretching. We ran several calculations with a variety
of initial conditions, including a wire attached to a top site of the rectangular 
tip, or to a bridge site, or to a hollow site. The
results showed that the changes between different starting points 
were significant, especially regarding the
value of the total elongation of the wire. Accordingly we decided to specialize
the simulation according to a single reasonable, but particular path. 
Starting from a perfect
geometry of wire and tips ($L= 25.6$\AA , 
which corresponds to the equilibrium bond length for each gold atoms in
the system), we compress the wire by progressively moving one tip closer to
the other (i.e. by decreasing the length L) and minimizing the total energy
of the system. The compression stops when the energy reaches a minimum 
(distance between fixed tips equal to $9.8$\AA).

At this point, we begin stretching the system to form a nanowire. These
two preparatory steps in the simulation allow us to generate a complete
\itshape{a priori} \normalfont geometry for all the moving atoms, a geometry 
which could and was normally different from the initial arrangement. In this way, 
we may obtain results that are less dependent on the initial configuration 
chosen for the system. 

The remaining problem concerns the choice on the number of moving atoms which participate
in the stretching process. We are mainly concerned by the electronic transport in the
nanowire and the tip geometry might appear a secondary problem.
However, it is important to check how the geometry of the whole 
tip-wire junction influences the conductance. That aspect
will be discussed in Sec. 3.

The elongation of the nanowire is simulated by slowly increasing
the distance between the two tips, avoiding abrupt variations
that could break the wire. For each stretching step, corresponding  to an
elongation of $0.02$\AA, the equilibrium configuration is determined
for all the moving atoms by minimizing the total energy.
That minimum corresponds to a constrained equilibrium configuration 
for the tip-suspended nanowire subject to a given stretching length. The two 
tips are then progressively moved apart until the wire breaks.

\subsection{Electronic transmission through a metallic nanowire}

To study the quantum transport of electrons through the nanowire, we adopt a very
simple one-electron model. The whole tip-nanowire-tip system is treated as a one-dimensional 
(1D) system, made up of perfect semi-infinite left tip and right leads, 
connecting through an extended defect,
formed by the monatomic nanowire the junctions and the first piece of the tips. Inside the perfect 
semi-infinite 1D leads, or reservoirs, electron propagation is described by a single-band 
tight binding Hamiltonian characterized by an on-site energy $E_{0}$ and the 
electron hopping integral $J$ between nearest neighbor sites.
  
\begin{figure}[ht]
\begin{center}
\includegraphics[width=400pt,height=150pt]{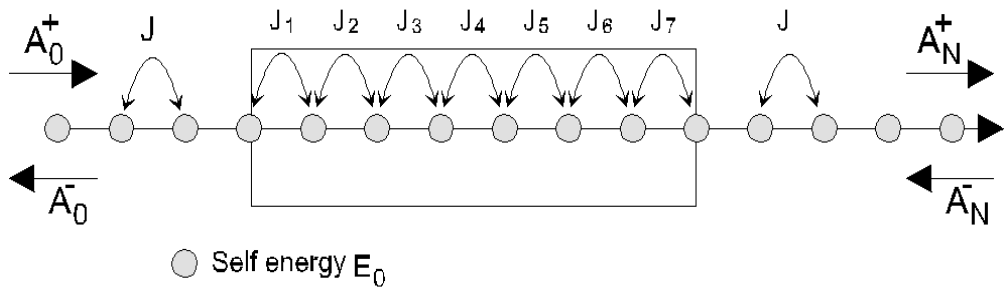}
\end{center}
\caption{One dimensional model for the calculation of the conductance of the stretched
nanowire connected to semi-infinite reservoirs.}
\label{fig:figure1}
\end{figure}


The defect, which corresponds to the nanowire, is assumed to consists simply
of a monatomic chain of $N$ atoms connected to the reservoirs. If and when  the distance 
between neighboring atoms fluctuates under stretching, the electron propagation
will be modified. To further simplify the treatment, we assume these positional fluctuations 
to affect the hopping integrals $J$, but not the on-site energies, retaining their fixed
value $E_0$. As a result, 
the electron propagation inside the nanowire will be characterized by a set of 
$N-1$ hopping integrals $J_{n}$, $n=1,..,N-1$. (see Fig.\ref{fig:figure1}) They can
be related to the distances and to the bulk integral $J$ according to the 
empirical form \cite{SPANJAARD} 
\begin{equation}
J_{n}=Jexp(\beta (1-r_{n}/r_{eq}))
\end{equation}
where $r_{n}$ stands for the distance between the ($n-1$)th and $n$th atoms
belonging to the nanowire and $r_{eq}$ denotes the equilibrium distance
(i.e. the lattice parameter of the perfect atomic chain $r_{eq}$=$a_0$=$2.49$\AA).
Note that when all the distances $r_{n}$ are set to $r_{eq}$, the different 
hopping integrals $J_{n}$ reduce to the bulk integral $J$ yielding an ideal system without any defects. 
While the model is clearly oversimplified in many aspects, it does serve our purpose
of pinpointing the effect of distance fluctuations on the conductance.

The Green's function formalism can be used to calculate the transmission
coefficient in this model. We consider an incident electron propagating freely from the
left tip with a wavevector $q$ and an energy $E=E_{0}+2Jcos(q)$. The electron 
is scattered by the nanowire leading to transmitted and reflected waves. The 
transmission coefficient is given by
\begin{equation}
t(E)=1+\frac{i}{2Jsin(q)}\sum_{n,m}T(n,m)e^{iq(x_{n}-x_{m})}
\end{equation}
where $x_{n}$ denotes the position of the $n$th atom of the 1D nanowire 
and where the T-matrix is a $N \times N$ matrix operator
\begin{equation}
T=V(1-G_{0}(E)V)^{-1}
\end{equation}
In Eq.(4), $G_{0}(E)$ denotes the Green's function operator of the ideal perfect 
infinite chain, whereas the operator $V$, known as the cleavage operator \cite{dob}, 
represents the perturbation caused by the fluctuations of the interatomic distances 
in the nanowire. This operator, whose dimension is again $N$, acts inside the $N$- atom
subspace.

The conductance $g$ through the region where scattering takes place is related
to the transmission coefficient of the electron via the Landauer formula \cite{landauer1,landauer2},
expressed at low temperature and low voltage, and for a single channel, as 
\begin{equation}
g=\frac{2e^{2}}{h}\mid t(E_{F})\mid ^{2}  \label{eq:landauer}
\end{equation}
where $E_{F}$ is the Fermi level of the electrons. Applying Eq.(\ref
{eq:landauer}) to our situation requires that the left and right hand sides
of the region where scattering takes place is connected to two reservoirs of 
electrons held at infinitesimally different electrochemical potentials. 
By assuming that the two semi-infinite chains play the role of reservoirs, 
our single-band model leads at half filling to a Fermi level $E_{F}$ equal to $E_{0}$.

\section{Results and discussion}

Fig. 2 displays typical snapshots during the formation and the stretching of
the wire. At the beginning of the simulation (Fig. 2a), all atoms in the wire and
tips are at their equilibrium distance $a_{0}=2.49$\AA  and 
$a=2.88$\AA, respectively. The ideal nanowire contains $7$ atoms. Compression proceeds 
until the two tips are stuck together and the total energy is a local 
minimum with nearest neighbor distance between Au atoms approximately equal 
to $a=2.80$\AA (Fig. 2b). At
this point, the initial configuration is lost and the moving part of the system is disordered :
the stretching process can start. Upon initial stretching, gold atoms tend to
arrange themselves in twisted chains until one atom is extracted from a tip (Fig. 2c).
Subsequently, a monatomic chain is formed by successive incorporation of
atoms from the tips into the wire (Fig. 2d). At a total length
corresponding to 8 atoms long (one more than the starting wire), the
nanowire breaks (Fig. 2e). These results obtained with the SMA potential are fully consistent with
those found using an effective medium potential \cite{RUBIO-BOLLINGER}. Note
that, as already mentioned in a previous paper \cite{PICAUDTOS},
incorporation of each new atom in the chain is obtained after an elongation $\Delta L$ of $L$
nearly equal to 1.5 \AA . This incorporation induces a reorganization of the wire,
particularly by decreasing the bond lengths between the gold atoms. 
This model reproduces quite well the limit distance from
which a nanowire can be elastically stretched  without dramatic change (1 
\AA\ in the experiments \cite{AGRAIT}). Between two successive incorporations, the values of 
the distances between the atoms in the wire and between the wire extremities and the apex of the tips
generally display a non uniform distribution. The stretching tends to increase 
the bond lengths in the wire, the disturbance being stronger on those atomic bonds 
close to the wire ends than on the bonds in the wire center.
In fact, we generally observe that the bonds linking the wire to the tips are weaker than the 
bonds between atoms of the wire. That also explains why a wire can be 
pulled out of tips and stretched out to such remarkable lengths without
breaking in the middle.

\begin{figure}[ht]
\begin{center}
\rotatebox{-90}{
\includegraphics[width=300pt,height=300pt]{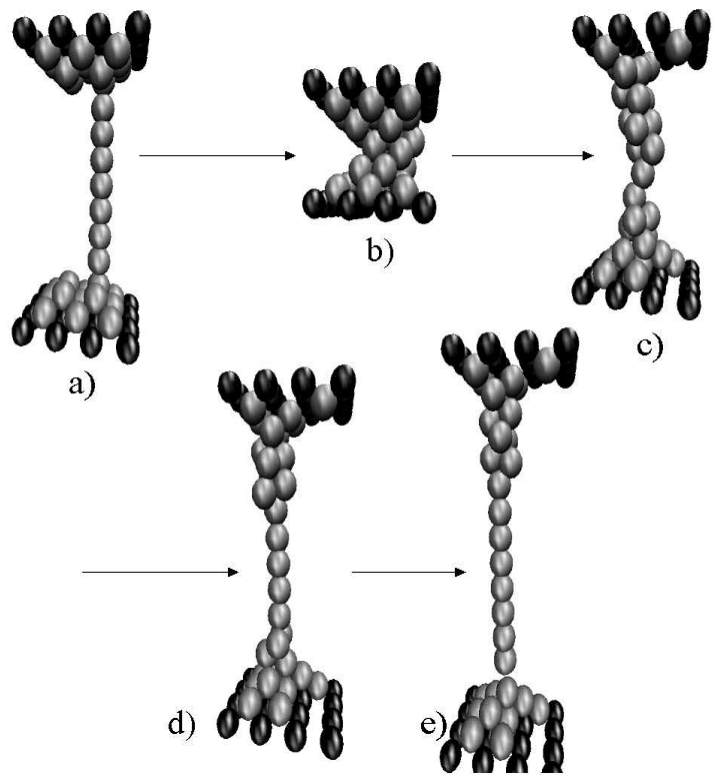}}
\end{center}
\caption{Mechanical model.  a) Starting configuration of the simulation. b)
the two tips are approached to the closest position; c) stretching has begun, and 
a single atom has been pulled out of the neck, initiating the monatomic
nanowire. This is also the starting point of conductance calculations
d) strained monatomic nanowire now reaching a 5- atoms length; e) maximum
elongation to an 8-atom long nanowire. At this point, the wire
breaks and conductance drops to zero. The total energy of the system is 
calculated with Gamba's semi-empirical potential \cite{GAMBA} and is minimized 
throughout during the stretching process.}
\label{fig:figure2}
\end{figure}

The changes of interatomic distances in the nanowire $r_{n}$ are extracted
from the simulation and turned into one dimensional effective distances  $x_{n}$ 
for conductance calculations. The  $x_{n}$ are the same as $r_{n}$ except at the
wire-tip junctions. 
There we assume conduction to proceed always via the
smallest bond length.

For the energy of the electrons generating the conduction in the nanocontact, we consider an
energy $E$ equal to and close to the Fermi level energy $E_{F}$ of the bulk gold. 

Fig. 3 shows the calculated behavior of the conductance of the total model
nanocontact of Fig. 2 during stretching, and represents our main result.

The conductance is close to $g_0$ as it should for a single channel,
the broad electronic band suffering only minor scattering from the weak 
perturbation represented by the distance fluctuations. The main characteristics 
we wish to focus upon here is precisely the conductance fluctuations 
that occur during stretching of the wire. 
These variations represent less than $5\%$ of $g_{0}$ and are
directly comparable with the value obtained in experiment \cite{AGRAIT}. 
Besides the 5\% magnitude, another point of similarity of the conductance curves 
in Fig. 3 and Fig. 5 with the experimental curves under stretching is the sharp slope
at the jumps. incorporation of an atom or by configurational change of the wire to its
extremities can interpret the constant distance ($1.5\AA $) between the conductance maxima. 

The fluctuations occur in the form of rather sharp conductance jumps with minima and maxima 
that depend strongly on the wire stretching. These jumps are directly
related to the fluctuations of the atomic bond distances in the nanocontact 
perturbing the electronic transmission inside the nanowire. If the nanowire length 
were infinite, the fluctuations, no matter how weak, would of course block 
conductance completely, and lead to an Anderson insulator\cite{ANDERSON}.
The inverse Anderson localization length depends on disorder, roughly proportional 
to the square of the hopping fluctuations. In a nanowire of finite and short 
fixed length like we have here, the conductance simply shows small fluctuating drops from $g_0$.
In general, conductance should drop exponentially with the inverse 
localization length. The exponential amplification expected of the effect on conductance
of hopping disorder may explains the abruptness of fluctuations seen in Fig. 3.

\begin{figure}[ht]
\begin{center}
\rotatebox{-90}{
\includegraphics[width=300pt,height=300pt]{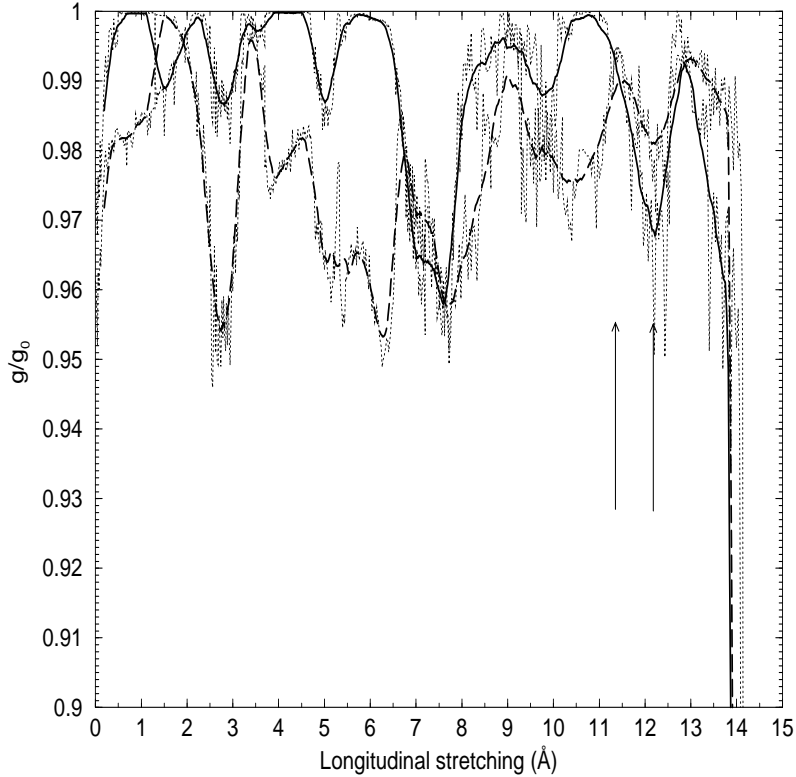}}
\end{center}
\caption{Behavior of the reduced conductance \itshape vs \normalfont
the wire stretching (dotted curves) for two incident electron energies. 
Full curve represents a spline fit of the conductance
fluctations during stretching of the wire for an incident energy
corresponding exactly to $E_F$. Broken curve represents a spline of the conductance
fluctations during stretching of the wire for an incident energy
corresponding to $E_{F}+J$.}. Arrows correspond to maxima and minima for 
the nanowire containing 7 atoms (see Fig. 4)
\label{fig:figure3}
\end{figure}

At finite voltage $W = \mu_L - \mu_R$, Landauer's ballistic conductance
is the same as in Eq. 5, only with $|t(E_F)|^2$ replaced by an average
of transmission  $|t(E)|^2 $ between $\mu_L$  and $\mu_R$.
We study in Fig. 3 the influence on transmission of the incident electronic energy 
by drawing $g$ for two distinct values of this energy (respectively equal to $E_{F}$  
and $E_{F}+J$). Although the conductance presents globally a similar jagged profile, 
a detailed examination of the curves shows some differences. A first difference
concerns the average amplitude of these variations. For $E$ = $E_{F}$
the mean curve tends in fact to remain localized around 0.99 $g_{o}$ while 
for energy $E_{F}+J$ the fluctuations take place around 0.975 $g_{o}$. 

This is relevant, for it implies a slight drop of {\em ballistic} 
conductance for finite and increasing voltage. The calculated decrease 
in fact corresponds to an increased tendency to Anderson localization as one 
moves away either sides from the center of the 1D band, the band being
centered precisely at $E_F$, where the localization length is maximum. 
Because of that, the resulting predicted conductance 
decrease is monotonic and relatively {\em uniform} for increasing voltage, that is for increasing
deviation of the electrochemical potential from $E_F$.

Interestingly, the experimentally observed conductance indeed does 
show a slight decrease with increasing voltage $W$\cite{AGRAIT}. However
the part of the conductance decrease that is uniform with voltage
(i.e., that has a roughly constant voltage derivative) is not major.
The major part is instead strongly peaked around a specific voltage,
ranging from 18 to 12 meV depending on the strain. This peak voltage in fact
corresponds to a longitudinal phonon frequency in the nanowire\cite{AGRAIT,PICAUDTOS},
and is evidence for the onset of non-ballistic, dissipative resistance 
in the nanowire.

Returning to Fig. 3 we note that in all cases the conduction minimum is around 0.95 $g_{0}$ 
in nice agreement with the experimental data. The maxima of all curves correspond
to the situation where a new atom is freshly incorporated in the wire thus permitting a general
compression of the chain. The minima mark the limit stretch just before the atom
pullout with incorporation and reorganization of the wire. 

To correlate more quantitatively the conductance behavior with the atomic bond 
length fluctuations, we have drawn in Fig. 4 the distribution of the bond lengths in two
wires formed by 7 atoms at the corresponding maxima (just after atom incorporation) and 
minima (just before atom incorporation). We see that this distribution , fitted by Gaussian 
profiles, broadens by $40\%$ during the stretching between two successive incorporation, and the
maximum of the profiles shifts from $2.48 \AA$ to $2.51 \AA$. This feature is quite general for 
long wires, and for instance we obtained for a 6 long wire the same identical
type of bond lengths distribution.

\begin{figure}[ht]
\begin{center}
\rotatebox{-90}{
\includegraphics[width=300pt,height=300pt]{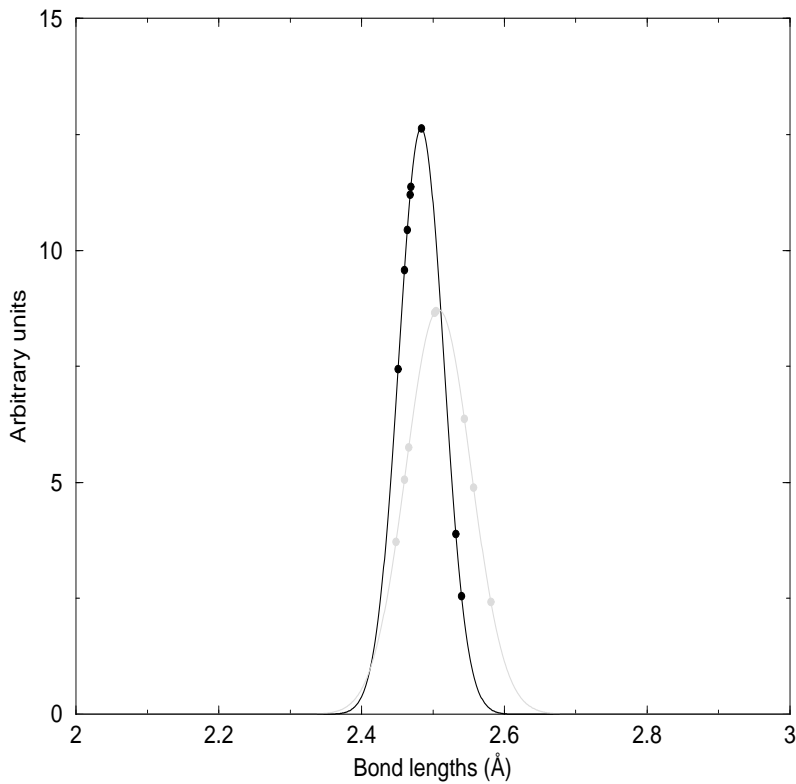}}
\end{center}
\caption{Bond length distribution between nearest neighbor atoms in the 7-atom 
monatomic nanowire at the maxima (dark curve) and minima (gray curve) of 
conductance of Fig.3}
\label{fig:figure4}
\end{figure}

We also mark an energy dependence of conductance connected to the wire length. 
When the nanowire is shorter than $8\AA$ (i.e., 4 atoms) the amplitude of the fluctuations are much
larger for $E=E_{F}+J$, so the shape of the two curves appear less similar than for longer nanowires
(7 or 8 atoms). Clearly, the stretching perturbs much more substantially the atomic 
bonds in the junctions than those amid the wire. The fluctuations are largely if not exclusively
a junction effect. As such, they are more important for a shorter nanowire.

Fig. 5 displays further evidence of the influence of the junctions between the nanowire 
and the two tips. In these calculations, we changed the bond length describing 
the tip-wire junction, all the time assuming the current-carrying path to remain
strictly one dimensional. The results show that while such a change does not modify
the shape of the conductance \itshape{vs} \normalfont the wire length, it does 
affect the mean conductance. Stretching that effective bond further and further 
away from its ideal value leads to a considerable conductance decrease.

\begin{figure}[ht]
\begin{center}
\includegraphics[width=300pt,height=300pt]{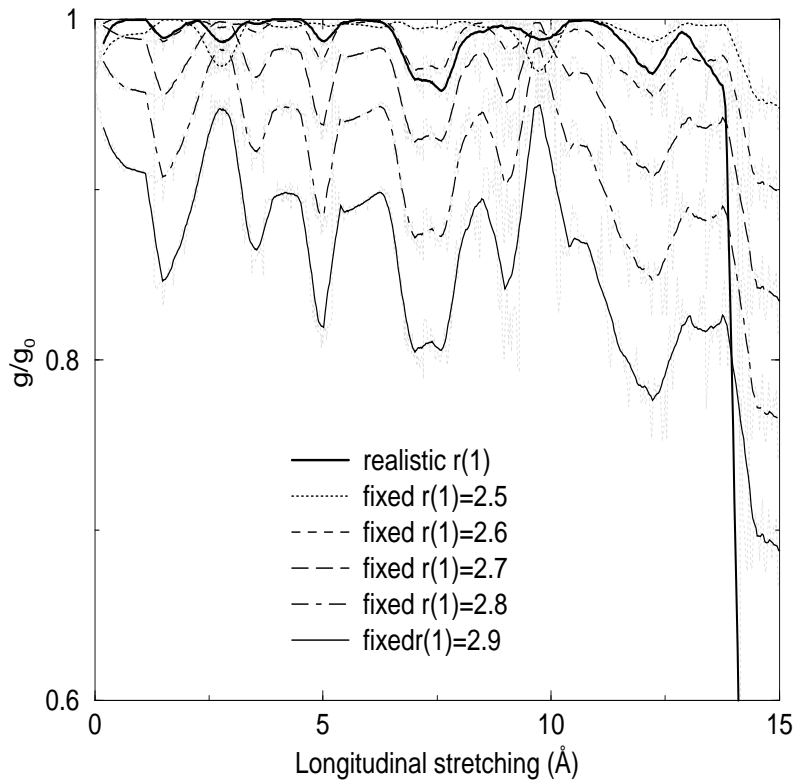}
\end{center}
\caption{Influence of the effective bond length representing the junction
between the nanowire and the tips during stretching.}
\label{fig:figure5}
\end{figure}

\section{Conclusions}

Motivated by recent experimental observations
showing that stretched monatomic gold wire between two tips presents systematic
conductance  fluctuations below $g_0$,
we carried out simulations of the behavior of electronic conductance during
the nanowire stretching up to the rupture of the contact. We find that the
observed 5\% conductance
fluctuations around $g_0$ can be explained as the result of continuously distributed
atomic bond lengths inside the stretched nanowire. The maxima of conductance are correlated to an 
atom incorporation in the wire which induces a compression of the chain while the minima correspond
to the maximum wire stretching, just before incorporation and/or reorganization.

\section {Acknowledgments}

Work done in SISSA was supported by MIUR through FIRB RBAU01LX5H,  FIRB RBAU017S8R, 
and through COFIN; by INFM;  and by the European Commission, through a fellowship to one of us
(F.P.) under contract ERBFMRXCT970155 (TMR FULPROP); and by CINECA. We
thank C. Untiedt for informing us of his results.

\newpage

\end{document}